\documentstyle{article}
\begin{document}
\def\Lemaitre{Lema\^\i{}tre}
\def\Painleve{Painlev\'e}
\title{Acoustic black holes: horizons, ergospheres,\\
       and Hawking radiation}
\author{Matt Visser\cite{e-mail}\\
        Physics Department\\
	Washington University\\
	Saint Louis\\
        Missouri 63130-4899\\
	USA}
\date{1 December 1997}
\maketitle
\centerline{gr-qc/9712010}
\begin{abstract}
It is a deceptively simple question to ask how acoustic disturbances
propagate in a non-homogeneous flowing fluid. Subject to suitable
restrictions, this question can be answered by invoking the language
of Lorentzian differential geometry. I begin this paper with a
pedagogical derivation of the following result: If the fluid is
barotropic and inviscid, and the flow is irrotational (though possibly
time dependent), then the equation of motion for the velocity
potential describing a sound wave is identical to that for a minimally
coupled massless scalar field propagating in a $(3+1)$--dimensional
Lorentzian geometry
\[ 
\Delta \psi \equiv 
{1\over\sqrt{-g}} \partial_\mu 
\left( \sqrt{-g}\; g^{\mu\nu} \; \partial_\nu\psi \right) = 
0.  
\]
The {\em acoustic metric} $g_{\mu\nu}(t,\vec x)$ governing the
propagation of sound depends algebraically on the density, flow
velocity, and local speed of sound. Even though the underlying
fluid dynamics is Newtonian, non-relativistic, and takes place in
flat space plus time, the fluctuations (sound waves) are governed
by an effective $(3+1)$--dimensional Lorentzian space-time geometry.
This rather simple physical system exhibits a remarkable connection
between classical Newtonian physics and the differential geometry
of curved $(3+1)$--dimensional Lorentzian spacetimes, and is the
basis underlying a deep and fruitful analogy between the black
holes of Einstein gravity and supersonic fluid flows. Many results
and definitions can be carried over directly from one system to
another. For example, I will show how to define the ergosphere,
trapped regions, acoustic apparent horizon, and acoustic event
horizon for a supersonic fluid flow, and will exhibit the close
relationship between the acoustic metric for the fluid flow
surrounding a point sink and the \Painleve--Gullstrand form of the
Schwarzschild metric for a black hole.  This analysis can be used
either to provide a concrete non-relativistic analogy for black
hole physics, or to provide a framework for attacking acoustics
problems with the full power of Lorentzian differential geometry.
\end{abstract}

\section{Introduction}

In 1981 Unruh developed a way of mapping certain aspects of black hole
physics into problems in the theory of supersonic acoustic
flows~\cite{Unruh81}. The connection between these two seemingly
disparate systems is both surprising and powerful, and has been
independently rediscovered several times over the ensuing decade and a
half~\cite{Visser93}. Over the last six years, a respectable
body of work has been developed using this analogy to investigate
micro-physical models that might underly the Hawking radiation process
from black holes (or acoustic holes --- ``dumb holes''), and to
investigate the extent to which the Hawking radiation process may be
independent of the physics of extremely high-energy trans--Planckian
modes~\cite{Jacobson91,Jacobson93,Unruh94,Brout,Jacobson95,Jacobson96,%
Corley-Jacobson96,Corley-Jacobson97,Corley97a,Corley97b,Reznik96,Reznik97}.

In this paper, I wish to take another look at the derivation of
the relationship between curved spacetimes and acoustics in flowing
fluids, to provide a pedagogically clear and precise derivation
using a minimum of technical assumptions, and to develop the analogy
somewhat further in directions not previously envisaged. In
particular, I will show how to define the notions of ergo-region,
trapped regions, acoustic apparent horizons, and acoustic event
horizons (both past and future) for supersonic fluid flows, and
show that in general it is necessary to keep these notions distinct.

As a particular example of a simple model exhibiting such behaviour I
write down the acoustic metric appropriate to a draining bathtub
[$(2+1)$ dimensions], and the equivalent vortex filament sink [$(3+1)$
dimensions].

I shall further show that the relationship between the Schwarzschild
geometry and the acoustic metric is clearest when the Schwarzschild
metric is written in the \Painleve--Gullstrand\footnote{This is
also often called the \Lemaitre\ form of the Schwarzschild metric.}
form~\cite{Painleve,Gullstrand,Lemaitre,Israel,Kraus-Wilczek}, and
that while the relationship is very close it is not exact. (It is
in fact impossible to obtain an acoustic metric that is {\em
identical} to the Schwarzschild metric, the best that one can
achieve is to obtain an acoustic metric that is conformally related
to the Schwarzschild metric.) If all one is interested in is either
the Hawking temperature or the behaviour in the immediate region
of the event horizon then the analogy is much closer in that the
conformal factor can be neglected.

For an arbitrary steady flow the ``surface gravity'' ({\em mutatis
mutandis}, the the Hawking temperature) of an acoustic horizon will
be shown to be proportional to a combination of the normal derivative
of the local speed of sound and the normal derivative of the normal
component of the fluid velocity at the horizon. In general, the
``surface gravity'' is
\begin{equation}
g_H = 
{1\over2} \; {\partial(c^2-v_\perp^2)\over \partial n} =
 c \; {\partial(c-v_\perp)\over\partial n}.
\end{equation}
(This generalizes the result of Unruh~\cite{Unruh81,Unruh94} to
the case where the speed of sound is position dependent and/or the
acoustic horizon is not the null surface of the time translation
Killing vector. This result is also compatible with that deduced
for the solid-state black holes of Reznik~\cite{Reznik97}, and with
the ``dirty black holes'' of~\cite{Visser92}.)

Finally, I shall show how to formulate the notion of a static (as
opposed to merely stationary) acoustic metric and exhibit the
constraint that must be satisfied in order to put the acoustic metric
into Schwarzschild coordinates. I point out that while this is a
perfectly acceptable and correct mathematical step, and a perfectly
reasonable thing to do in general relativity, it is (I claim) a good
way to get confused when doing acoustics --- from the Newtonian view
underlying the equations of fluid motion the Schwarzschild coordinate
system corresponds to a very peculiar way of synchronizing (or rather,
de-synchronizing) your clocks.

To begin the discussion: I address the deceptively simple question of
how acoustic disturbances propagate in a non-homogeneous flowing
fluid.  It is well known that for a static homogeneous inviscid fluid
the propagation of sound waves is governed by the simple
equation~\cite{Lamb,Landau-Lifshitz,Milne-Thomson,Skudrzyk}
\begin{equation}
\partial_t^2 \psi = c^2 \nabla^2 \psi.
\end{equation}
(Here $c \equiv \hbox{speed of sound}$.) Generalizing this result to a
fluid that is non-homogeneous, or to a fluid that is in motion,
possibly even in non-steady motion, is more subtle than it at first
would appear.

An important aspect of this paper is to provide a pedagogical proof of
the following theorem:
\paragraph{Theorem}{\em
If a fluid is barotropic and inviscid, and the flow is irrotational
(though possibly time dependent) then the equation of motion for the
velocity potential describing an acoustic disturbance is identical to
the d'Alembertian equation of motion for a minimally coupled massless
scalar field propagating in a $(3+1)$--dimensional Lorentzian
geometry}
\begin{equation}
\Delta \psi \equiv 
{1\over\sqrt{-g}} 
\partial_\mu 
\left( \sqrt{-g} \; g^{\mu\nu} \; \partial_\nu \psi \right) = 0.
\end{equation}
Under these conditions, the propagation of sound is governed by an
{\em acoustic metric} --- $g_{\mu\nu}(t,\vec x)$. This acoustic
metric describes a $(3+1)$--dimensional Lorentzian (pseudo--Riemannian)
geometry. The metric depends algebraically on the density, velocity
of flow, and local speed of sound in the fluid. Specifically
\begin{equation}
g_{\mu\nu}(t,\vec x) 
\equiv {\rho\over c} 
\left[ \matrix{-(c^2-v^2)&\vdots&-{\vec v}\cr
               \cdots\cdots\cdots\cdots&\cdot&\cdots\cdots\cr
	       -{\vec v}&\vdots& I\cr } \right].
\end{equation}
(Here $I$ is the $3\times3$ identity matrix.) In general, when
the fluid is non-homogeneous and flowing, the {\em acoustic Riemann
tensor} associated with this Lorentzian metric will be nonzero.

It is quite remarkable that even though the underlying fluid dynamics
is Newtonian, nonrelativistic, and takes place in flat space plus
time, the fluctuations (sound waves) are governed by a curved
$(3+1)$--dimensional Lorentzian (pseudo-Riemannian) spacetime
geometry.

For practitioners of general relativity this paper describes a very
simple and concrete physical model for certain classes of Lorentzian
spacetimes, including black holes.  On the other hand, the discussion
of this paper is also potentially of interest to practitioners of
continuum mechanics and fluid dynamics in that it provides a simple
concrete introduction to Lorentzian differential geometric techniques.

\section{Fluid Dynamics}

\subsection{Fundamental equations}

The fundamental equations of fluid
dynamics~\cite{Lamb,Landau-Lifshitz,Milne-Thomson,Skudrzyk} are
the equation of continuity
\begin{equation}
\partial_t \rho + \nabla\cdot(\rho \; \vec v) = 0,
\label{E-continuity}
\end{equation}
and Euler's equation
\begin{equation}
\rho {d \vec v\over dt} \equiv 
\rho \left[ \partial_t \vec v + (\vec v \cdot \nabla) \vec v \right] =
\vec F.
\label{E-euler}
\end{equation}
I start the analysis by assuming the fluid to be inviscid (zero
viscosity), with the only forces present being those due to pressure,
plus Newtonian gravity, and  with the inclusion of any arbitrary
gradient-derived and possibly even time-dependent externally-imposed
body force. Then
\begin{equation}
\vec F = - \nabla p - \rho \nabla \phi - \rho \nabla\Phi.
\label{E-force}
\end{equation}
Here $\phi$ denotes the Newtonian gravitational potential, while
$\Phi$ denotes the potential of the external driving force (which may
in fact be zero).\footnote{These two terms are lumped together
without comment in~\cite{Unruh81}, and are neglected in~\cite{Unruh94}.}

Via standard manipulations the Euler equation can be rewritten as
\begin{equation}
\partial_t {\vec v} = 
\vec v \times ( \nabla \times \vec v)  - {1\over\rho} \nabla p
- \nabla\left( {{1\over2}} v^2 + \phi + \Phi\right).
\label{E-euler-2}
\end{equation}
Now take the flow to be {\em vorticity free}, that is, {\em locally
irrotational}.\footnote{The irrotational condition is automatically
satisfied for the superfluid component of physical superfluids.
This point has been emphasised by Comer~\cite{Comer}, who has also
pointed out that in superfluids there will be multiple acoustic
metrics (and multiple acoustic horizons) corresponding to first
and second sound. Even for normal fluids, vorticity free flows are
common, especially in situations of high symmetry.} Introduce the
velocity potential $\psi$ such that $\vec v = -\nabla \psi$, at
least locally.\footnote{It is sufficient that the flow be vorticity
free, $\vec\nabla\times\vec v=0$, so that velocity potentials exist
on an atlas of open patches --- this enables us to handle vortex
filaments, where the vorticity is concentrated into a thin vortex
core, provided we do not attempt to probe the vortex core itself.
I do not need $\psi$ to be globally defined.}  If one further takes
the fluid to be {\em barotropic}\footnote{An unstated assumption
of this type is implicit, though not explicit, in the analysis of
reference~\cite{Unruh81}.  On the other hand, reference~\cite{Unruh94}
explicitly makes the stronger assumption that the fluid is {\em
isentropic}.  (That is, the specific entropy density is taken to
be constant throughout the fluid.)  This is a stronger assumption
than is actually required, and the weaker barotropic assumption
used here is sufficient. In particular, the present derivation also
applies to isothermal perturbations of an isothermal fluid.} (this
means that $\rho$ is a function of $p$ only), it becomes possible
to define
\begin{equation}
h(p) = \int_0^p {dp'\over\rho(p')}; 
\qquad \hbox{so that} \qquad
\nabla h = {1\over\rho} \; \nabla p.
\end{equation}
Thus the specific enthalpy, $h(p)$, is a function of $p$ only.
Euler's equation now reduces to
\begin{equation}
-\partial_t \psi + h + {1\over2} (\nabla\psi)^2 
+ \phi  + \Phi = 0.
\label{E-bernoulli}
\end{equation}
This is a version of Bernoulli's equation in the presence of external
driving forces.

\subsection{Fluctuations}

Now linearize these equations of motion around some assumed background
$(\rho_0,p_0,\psi_0)$.  Set $\rho = \rho_0 + \epsilon \rho_1 +
O(\epsilon^2)$, $p = p_0 + \epsilon p_1 +O(\epsilon^2)$, and $\psi
= \psi_0 + \epsilon \psi_1 + O(\epsilon^2)$.  The gravitational
potential $\phi$, and driving potential $\Phi$, are taken to be
fixed and external\footnote{Fixed means that I do not allow
back-reaction to modify the gravitational or driving potentials.
Fixed does not necessarily mean time independent, as I explicitly
wish to allow the possibility of time dependent external driving
forces.}.  Sound is {\em defined} to be these linearized fluctuations
in the dynamical quantities.  Please note that this is the {\em
standard definition} of sound and more generally of acoustical
disturbances. In principle, of course, one is really interested in
solving the complete equations of motion for the fluid variables
$(\rho, p, \psi)$. In practice, it is both traditional and extremely
useful to separate the exact motion, described by the exact variables,
$(\rho, p, \psi)$, into some average bulk motion, $(\rho_0,p_0,\psi_0)$,
plus low amplitude acoustic disturbances, $(\epsilon \rho_1,\epsilon
p_1,\epsilon \psi_1)$. See, for
example~\cite{Lamb,Landau-Lifshitz,Milne-Thomson,Skudrzyk}.

Since this is a subtle issue that I have seen cause considerable
confusion in the past, let me be even more explicit by asking the
rhetorical question: {\em ``How can we tell the difference between a
wind gust and a sound wave?''} The answer is that the difference is to
some extent a matter of convention --- sufficiently low-frequency
long-wavelength disturbances (wind gusts) are conventionally lumped in
with the average bulk motion. Higher-frequency, shorter-wavelength
disturbances are conventionally described as acoustic disturbances. If
you wish to be hyper-technical, we can introduce a high-pass filter
function to define the bulk motion by suitably averaging the exact
fluid motion. There are no deep physical principles at stake here ---
merely an issue of convention.

The place where we are making a specific physical assumption that
restricts the validity of our analysis is in the requirement that the
amplitude of the high-frequency short-wavelength disturbances be
small. This is the assumption underlying the linearization programme,
and this is why sufficiently high-amplitude sound waves must be
treated by direct solution of the full equations of fluid dynamics.

Linearizing the continuity equation results in the pair of equations
\begin{eqnarray}
&&
\partial_t \rho_0 + 
\nabla\cdot(\rho_0 \; \vec v_0) = 0,
\\
&&
\partial_t \rho_1 + 
\nabla\cdot(\rho_1 \; \vec v_0 + \rho_0 \; \vec v_1) = 0.
\end{eqnarray}
Now, the barotropic condition implies
\begin{equation}
h(p) = 
h(p_0 + \epsilon p_1 +  O(\epsilon^2)) = 
h_0 + \epsilon \; {p_1\over\rho_0} + O(\epsilon^2).
\end{equation}
Use this result in linearizing the Euler equation. We obtain the pair
\begin{eqnarray}
&&-\partial_t \psi_0 + h_0 + {{1\over2}} (\nabla\psi_0)^2 
+ \phi + \Phi = 0.
\\
&&-\partial_t \psi_1 + {p_1\over\rho_0} - \vec v_0 \cdot \nabla\psi_1 = 0.
\end{eqnarray}
This last equation may be rearranged to yield
\begin{equation}
p_1 =  \rho_0 ( \partial_t \psi_1 + \vec v_0 \cdot \nabla\psi_1 ).
\label{E-linear-euler}
\end{equation}
Use the barotropic assumption to relate
\begin{equation}
\rho_1 = 
{\partial \rho\over\partial p} \; p_1 = 
{\partial \rho\over\partial p} \; \rho_0  \;
( \partial_t \psi_1 + \vec v_0 \cdot \nabla\psi_1 ). 
\label{E-linear-barotropic}
\end{equation}
Now substitute this consequence of the linearized Euler equation into
the linearized equation of continuity. We finally obtain, up to an
overall sign, the wave equation:
\begin{eqnarray}
&-& \partial_t  
     \left( {\partial\rho\over\partial p} \; \rho_0 \; 
            ( \partial_t \psi_1 + \vec v_0 \cdot \nabla\psi_1 ) 
     \right)
\nonumber\\
&+& \nabla \cdot 
     \left( \rho_0 \; \nabla\psi_1 
            - {\partial\rho\over\partial p} \; \rho_0 \; \vec v_0 \;
	      ( \partial_t \psi_1 + \vec v_0 \cdot \nabla\psi_1 ) 	    
     \right)
=0.
\label{E-wave-physical}
\end{eqnarray}
This wave equation describes the propagation of the linearized
scalar potential $\psi_1$. Once $\psi_1$ is determined, equation
(\ref{E-linear-euler}) determines $p_1$, and equation
(\ref{E-linear-barotropic}) then determines $\rho_1$.  Thus this
wave equation  completely determines the propagation of acoustic
disturbances.  The background fields $p_0$, $\rho_0$ and $\vec v_0
= - \nabla \psi_0$, which appear as time-dependent and
position-dependent coefficients in this wave equation, are
constrained to solve the equations of fluid motion for an
externally-driven, barotropic, inviscid, and irrotational flow.
Apart from these constraints, they are otherwise permitted to have
{\em arbitrary} temporal and spatial dependencies.

Now, written in this form, the physical import of this wave equation
is somewhat less than pellucid. To simplify things algebraically,
observe that the local speed of sound is defined by
\begin{equation}
c^{-2} \equiv {\partial\rho\over\partial p}. 
\end{equation}
Now construct the symmetric $4\times4$ matrix
\begin{equation}
f^{\mu\nu}(t,\vec x) \equiv 
{\rho_0\over c^2}
\left[ \matrix{-1&\vdots&-v_0^j\cr
               \cdots\cdots&\cdot&\cdots\cdots\cdots\cdots\cr
	       -v_0^i&\vdots&( c^2\delta^{ij} - v_0^i v_0^j )\cr } 
\right].
\label{E-explicit}	       
\end{equation}
(Greek indices run from $0$--$3$, while Roman indices run from
$1$--$3$.)  Then, introducing $(3+1)$--dimensional space-time
coordinates --- $x^\mu \equiv (t; x^i)$ --- the above wave equation
(\ref{E-wave-physical}) is easily rewritten as
\begin{equation}
\partial_\mu ( f^{\mu\nu} \; \partial_\nu \psi_1) = 0.
\end{equation}
This remarkably compact formulation is completely equivalent to
equation (\ref{E-wave-physical}) and is a much more promising
stepping-stone for further manipulations. The remaining steps are
a straightforward application of the techniques of curved space
$(3+1)$--dimensional Lorentzian geometry.

\section{Lorentzian Geometry}

In any Lorentzian (that is, pseudo--Riemannian) manifold the curved
space scalar d'Alembertian is given in terms of the metric
$g_{\mu\nu}(t,\vec x)$ by (see, for
example,~\cite{Fock,Moller,MTW,Hawking-Ellis,Wald})
\begin{equation}
\Delta \psi \equiv 
{1\over\sqrt{-g}} 
\partial_\mu \left( \sqrt{-g} \; g^{\mu\nu} \; \partial_\nu \psi \right).
\end{equation}
The inverse metric, $g^{\mu\nu}(t,\vec x)$, is pointwise the matrix
inverse of $g_{\mu\nu}(t,\vec x)$, while $g \equiv \det(g_{\mu\nu})$.
Thus one can rewrite the physically derived wave equation
(\ref{E-wave-physical}) in terms of the d'Alembertian provided one
identifies

\begin{equation}
\sqrt{-g} \; g^{\mu\nu} = f^{\mu\nu}.
\end{equation}
This implies, on the one hand 
\begin{equation}
\det(f^{\mu\nu}) = (\sqrt{-g})^4 \; g^{-1} = g.
\end{equation} 
On the other hand, from the explicit expression (\ref{E-explicit}),
expanding the determinant in minors
\begin{equation}
\det(f^{\mu\nu}) 
= 
\left({\rho_0\over c^2}\right)^4 \cdot
\left[(-1) \cdot (c^2 - v_0^2) - (-v_0)^2\right] \cdot
\left[c^2\right] \cdot
 \left[c^2\right]
=
- {\rho_0^4\over c^2}.
\end{equation} 
Thus
\begin{equation}
g = - {\rho_0^4\over c^2}; \qquad \sqrt{-g} = {\rho_0^2\over c}.
\end{equation}
We can therefore pick off the coefficients of the inverse acoustic
metric\footnote{There is a minor typo, a missing factor of $c$,
in~\cite{Unruh94}.}
\begin{equation}
g^{\mu\nu}(t,\vec x) \equiv 
{1\over \rho_0 c}
\left[ \matrix{-1&\vdots&-v_0^j\cr
               \cdots\cdots&\cdot&\cdots\cdots\cdots\cdots\cr
	       -v_0^i&\vdots&(c^2 \delta^{ij} - v_0^i v_0^j )\cr } 
\right].	       
\end{equation}
We could now determine the metric itself simply by inverting this
$4\times4$ matrix. On the other hand, it is even easier to recognize
that one has in front of one an example of the Arnowitt--Deser--Misner
split of a $(3+1)$--dimensional  Lorentzian spacetime metric into
space + time, more commonly used in discussing initial value data
in Einstein's theory of gravity --- general relativity. (See, for
example, \cite{MTW} pp 505--508.) The acoustic metric is
\begin{equation}
g_{\mu\nu} \equiv 
{\rho_0 \over  c}
\left[ \matrix{-(c^2-v_0^2)&\vdots&-v_0^j\cr
               \cdots\cdots\cdots\cdots&\cdot&\cdots\cdots\cr
	       -v_0^i&\vdots&\delta_{ij}\cr } 
\right].	       
\end{equation}
Equivalently, the acoustic interval can be expressed as
\begin{equation}
ds^2 \equiv g_{\mu\nu} \; dx^\mu \; dx^\nu =
{\rho_0\over c} 
\left[
- c^2 dt^2 + (dx^i - v_0^i \; dt) \; \delta_{ij} \; (dx^j - v_0^j \; dt )
\right].
\end{equation}
A few brief comments should be made before proceeding:
\begin{itemize}
\item
Observe that the signature of this metric is indeed $(-,+,+,+)$, as
it should be to be regarded as Lorentzian.
\item
It should be emphasized that there are two distinct metrics relevant
to the current discussion:
\begin{itemize}
\item
The {\em physical spacetime metric} is just the usual flat metric
of Minkowski space
\begin{equation}
\eta_{\mu\nu} \equiv 
({\rm diag}[-c_{light}^2,1,1,1])_{\mu\nu}. 
\end{equation}
(Here $c_{light} = \hbox{speed of light}$.) The fluid particles couple
only to the physical metric $\eta_{\mu\nu}$. In fact the fluid
motion is completely non--relativistic --- $||v_0|| \ll c_{light}$.
\item
Sound waves on the other hand, do not ``see'' the physical metric
at all. Acoustic perturbations couple only to the {\em acoustic
metric} $g_{\mu\nu}$.
\end{itemize}
The geometry determined by the acoustic metric does however inherit
some key properties from the existence of the underlying flat
physical metric.
\item
For instance, the topology of the manifold does not depend on the
particular metric considered.  The acoustic geometry inherits
the underlying topology of the physical metric --- $\Re^4$ --- with
possibly a few regions excised (due to imposed boundary conditions).
\item
Furthermore, the acoustic geometry  automatically inherits the
property of ``stable causality''~\cite{Hawking-Ellis,Wald}. Note that
\begin{equation}
g^{\mu\nu} \, (\nabla_\mu t) \, (\nabla_\nu t) = -{1\over\rho_0 c} < 0.
\end{equation}
This precludes some of the more entertaining causality-related
pathologies that sometimes arise in general relativity. (For a
discussion of causal pathologies, see for example~\cite{Visser95}).
\item
Other concepts that translate immediately are those of ``ergo-region'',
``trapped surface'', ``apparent horizon'', and ``event horizon''.
These notions will be developed fully in the following section.
\item
The properly normalized four-velocity of the fluid is
\begin{equation}
V^\mu = {(1;v^i_0)\over\sqrt{\rho_0 \; c}}.
\end{equation}
This is related to the gradient of the natural time parameter by
\begin{equation}
\nabla_\mu t = (1,0,0,0); \qquad \qquad 
\nabla^\mu t =  -{(1;v^i_0)\over\rho_0 \; c} 
             = -{V^\mu\over\sqrt{\rho_0 \; c}}.
\end{equation}
Thus the integral curves of the fluid velocity field are orthogonal
(in the Lorentzian metric) to the constant time surfaces. The
acoustic proper time along the fluid flow lines (streamlines) is
\begin{equation}
\tau = \int \sqrt{\rho_0 \; c} \; dt,
\end{equation}
and the integral curves are geodesics of the acoustic metric if
and only if $\rho_0 c$ is position independent.
\item
Observe that in a completely general $(3+1)$--dimensional Lorentzian
geometry the metric has 6 degrees of freedom per point in spacetime.
($4\times4$ symmetric matrix $\Rightarrow$ $10$ independent
components; then subtract $4$ coordinate conditions).  In contrast,
the acoustic metric is more constrained. Being specified completely by
the three scalars $\psi_0(t, \vec x)$, $\rho_0(t, \vec x)$, and $c(t,
\vec x)$, the acoustic metric has at most $3$ degrees of freedom per
point in spacetime. The equation of continuity actually reduces this to
$2$ degrees of freedom which can be taken to be $\psi_0(t, \vec x)$
and $c(t, \vec x)$.
\item
A point of notation: Where the general relativist uses the word
``stationary'' the fluid dynamicist uses the phrase ``steady flow''.
The general-relativistic word ``static'' translates to a rather messy
constraint on the fluid flow (to be discussed more fully below).
\item
Finally, I should add that in Einstein gravity the spacetime metric is
related to the distribution of matter by the non-linear
Einstein--Hilbert differential equations. In contrast, in the present
context, the acoustic metric is related to the distribution of matter
in a simple algebraic fashion.
\end{itemize}

\section{Ergo-regions, trapped surfaces, and acoustic horizons}

Let's start with the notion of an ergo-region: Consider integral
curves of the vector $ K^\mu \equiv (\partial/\partial t)^\mu =
(1,0,0,0)^\mu $.  (If the flow is steady then this is the time
translation Killing vector.  Even if the flow is not steady the
background Minkowski metric provides us with a natural definition of
``at rest''.)  Then\footnote{Henceforth, in the interests of
notational simplicity, I shall drop the explicit subscript $0$ on
background field quantities unless there is risk of confusion.}
\begin{equation}
g_{\mu\nu} \; (\partial/\partial t)^\mu \;(\partial/\partial t)^\nu = 
g_{tt} = 
-[c^2 - v^2]. 
\end{equation}
This changes sign when $ ||\vec v|| > c $.  Thus any region of
supersonic flow is an ergo-region.  (And the boundary of the
ergo-region may be deemed to be the ergosphere.) The analogue of this
behaviour in general relativity is the ergosphere surrounding any
spinning black hole --- it is a region where space ``moves'' with
superluminal velocity relative to the fixed
stars~\cite{MTW,Hawking-Ellis,Wald}.

A trapped surface in acoustics is defined as follows: take any
closed two-surface. If the fluid velocity is everywhere inward-pointing
and the normal component of the fluid velocity is everywhere greater
than the local speed of sound, then no matter what direction a
sound wave propagates, it will be swept inward by the fluid flow
and be trapped inside the surface. The surface is then said to be
outer-trapped. (For comparison with the usual situation in general
relativity see \cite[pages 319-323]{Hawking-Ellis} or \cite[pages
310--311]{Wald}.) Inner-trapped surfaces (anti-trapped surfaces)
can be defined by demanding that the fluid flow is everywhere
outward-pointing with supersonic normal component. It is only
because of the fact that the background Minkowski metric provides
a natural definition of ``at rest'' that we can adopt such a simple
definition. In ordinary general relativity we need to develop
additional machinery, such as the notion of the ``expansion'' of
bundles of ingoing and outgoing null geodesics, before defining
trapped surfaces --- that the above definition is equivalent to
the usual one follows from the discussion on pages 262--263 of
Hawking and Ellis \cite{Hawking-Ellis}.  The acoustic trapped region
is now defined as the region containing outer trapped surfaces,
and the acoustic (future) apparent horizon as the boundary of the
trapped region. (We can also define anti-trapped regions and past
apparent horizons but these notions are of limited utility in
general relativity.)

The event horizon (absolute horizon) is defined, as in general
relativity, by demanding that it be the boundary of the region from
which null geodesics (phonons) cannot escape.  This is actually
the future event horizon. A past event horizon can be defined in
terms of the boundary of the region that cannot be reached by
incoming phonons --- strictly speaking this requires us to define
notions of past and future null infinities, but I will simply take
all relevant incantations as understood. In particular the event
horizon is a null surface, the generators of which are null geodesics.

In all stationary geometries the apparent and event horizons
coincide, and the distinction is immaterial. In time-dependent
geometries the distinction is often important.  When computing the
surface gravity I shall restrict attention to stationary geometries
(steady flow). In fluid flows of high symmetry, (spherical symmetry,
plane symmetry) the ergosphere may coincide with the acoustic
apparent horizon, or even the acoustic event horizon. This is the
analogue of the result in general relativity that for static (as
opposed to stationary) black holes the ergosphere and event horizon
coincide.

\section{Vortex geometries}

As an example of a fluid flow where the distinction between ergosphere
and acoustic event horizon is critical consider the ``draining
bathtub'' fluid flow. I model a draining bathtub by a $(2+1)$
dimensional flow with a sink at the origin. The equation of continuity
implies that for the radial component of the fluid velocity we must
have
\begin{equation}
\rho \; v^{\hat r} \propto {1\over r}.
\end{equation}
In the tangential direction, the requirement that the flow be
vorticity free (apart from a possible delta-function contribution at
the vortex core) implies, via Stokes' theorem, that 
\begin{equation}
v^{\hat t} \propto  {1\over r}.
\end{equation}
On the other hand, assuming conservation of angular momentum (this
places a constraint on the external body forces by assuming the absence
of external torques) implies the slightly different constraint
\begin{equation}
\rho  \; v^{\hat t} \propto  {1\over r}.
\end{equation}
Combining these constraints, the background density $\rho$ must be
constant (position-independent) throughout the flow (which
automatically implies that the background pressure $p$ and speed of
sound $c$ are also constant throughout the fluid flow). Furthermore
for the background velocity potential we must then have
\begin{equation}
\psi(r,\theta) = A\; \ln(r/a) + B \;\theta.
\end{equation}
Note that, as we have previously hinted, the velocity potential is not a
true function (because it has a discontinuity on going through $2\pi$
radians). The velocity potential must be interpreted as being defined
patch-wise on overlapping regions surrounding the vortex core at
$r=0$. The velocity of the fluid flow is
\begin{equation}
\vec v = {(A \; \hat r + B \; \hat\theta)\over r}.
\end{equation}
Dropping a position-independent prefactor, the acoustic metric for a draining
bathtub is explicitly given by
\begin{equation}
ds^2 = 
- c^2 dt^2 
+ \left(dr - {A\over r} dt\right)^2 
+ \left(r \, d\theta - {B\over r} dt\right)^2.
\end{equation}
Equivalently
\begin{equation}
ds^2 = 
- \left(c^2 -{A^2+B^2\over r^2}\right) dt^2   
- 2{A\over r} \, dr \, dt  - 2 B \, d\theta \, dt + dr^2 + r^2 d\theta^2.
\end{equation}
A similiar metric, restricted to $A=0$ (no radial flow), and
generalized to an anisotropic speed of sound, has been exhibited by
Volovik~\cite{Volovik}, that metric being a model for the acoustic
geometry surrounding physical vortices in superfluid ${}^3$He. (For a
survey of the many analogies and similarities between the physics of
superfluid ${}3$He and the Standard Electroweak Model
see~\cite{Volovik-Vachaspati}, this reference is also useful as
background to understanding the Lorentzian geometric aspects of
${}^3$He fluid flow.) Note that the metric given above is {\em not}
identical to the metric of a spinning cosmic string, which would
instead take the form~\cite{Visser95}
\begin{equation}
ds^2 = 
- c^2 (dt - {\tilde A} \, d\theta)^2  + dr^2 + (1-{\tilde B}) r^2 d\theta^2.
\end{equation}
In conformity with previous comments, the vortex fluid flow is seen to
possess an acoustic metric that is stably causal and which does not
involve closed timelike curves. (At large distances it is possible to
{\em approximate} the vortex geometry by a spinning cosmic
string~\cite{Volovik}, but this approximation becomes progressively
worse as the core is approached.)

The ergosphere forms at
\begin{equation}
r_{ergosphere} = {\sqrt{A^2 + B^2}\over c}.
\end{equation}
Note that the sign of $A$ is irrelevant in defining the ergosphere and
ergo-region: it does not matter if the vortex core is a source or a
sink.

The acoustic event horizon forms once the radial component of the
fluid velocity exceeds the speed of sound, that is at
\begin{equation}
r_{horizon} = {|A|\over c}.
\end{equation}
The sign of $A$ now makes a difference. For $A<0$ we are dealing with
a future acoustic horizon (acoustic black hole), while for $A>0$ we
are dealing with a past event horizon (acoustic white hole).

Though this construction has been phrased in $(2+1)$ dimensions we are
of course free to add an extra dimension by going to $(3+1)$
dimensions and interpreting the result as a superposition of an
ordinary vortex filament and a line source (or line sink).
\begin{equation}
ds^2 = 
- c^2 dt^2 
+ \left(dr - {A\over r} dt\right)^2 
+ \left(r \, d\theta - {B\over r} dt\right)^2 + dz^2.
\end{equation}
%

\section{Slab geometries}

A popular model for the investigation of event horizons in the
acoustic analogy is the one-dimensional slab geometry where the
velocity is always along the $z$ direction and the velocity profile
depends only on $z$. The continuity equation then implies that
$\rho(z) v(z)$ is a constant, and the acoustic metric becomes
\begin{equation}
ds^2  \propto  {1\over v(z) c(z) }
\left[ 
- c(z)^2 dt^2 + \left\{ dz - v(z) dt \right\}^2 + dx^2 + dy^2 
\right].
\end{equation}
That is
\begin{equation}
ds^2  \propto  {1\over v(z) c(z) }
\left[ 
- \left\{c(z)^2-v(z)^2\right\}  dt^2 - 2 v(z) dz dt+ dx^2 + dy^2 + dz^2
\right].
\end{equation}
If we set $c=1$ and ignore the conformal factor we have the toy
model acoustic geometry discussed by Unruh~\cite[page 2828, equation
(8)]{Unruh94} Jacobson~\cite[page 7085, equation (4)]{Jacobson96},
Corley and Jacobson~\cite{Corley-Jacobson96}, and Corley~\cite{Corley97a}.
(Since the conformal factor is regular at the event horizon, we
know that the surface gravity and Hawking temperature are independent
of this conformal factor~\cite{Jacobson93b}.) In the general case
it is important to realise that the flow can go supersonic for
either of two reasons: the fluid could speed up, or the speed of
sound could decrease.  When it comes to calculating the ``surface
gravity'' both of these effects will have to be taken into account.

\section{The \Painleve--Gullstrand line element}

To see how close the acoustic metric can get to reproducing the
Schwarzschild geometry it is first useful to introduce one of the
more exotic representations of the Schwarzschild geometry: the
\Painleve--Gullstrand line element, which is simply an unusual
choice of coordinates on the Schwarzschild spacetime.\footnote{The
\Painleve--Gullstrand line element is often called the \Lemaitre\ 
line element.} In modern notation the Schwarzschild geometry in
ingoing ($+$) and outgoing ($-$) \Painleve--Gullstrand coordinates
may be written as:
\begin{equation}
ds^2 = 
- dt^2 +
\left( dr \pm \sqrt{2GM\over r} dt \right)^2 
+ r^2\left( d\theta^2 + \sin^2\theta \; d\phi^2 \right).
\end{equation}
Equivalently
\begin{equation}
ds^2 = 
- \left(1-{2GM\over r}\right) dt^2 
\pm \sqrt{2GM\over r}\, dr \, dt 
+ dr^2 + r^2\left( d\theta^2 + \sin^2\theta \; d\phi^2 \right).
\end{equation}
This representation of the Schwarzschild geometry is not
particularly well-known and has been rediscovered several times this
century. See for instance \Painleve~\cite{Painleve},
Gullstrand~\cite{Gullstrand}, \Lemaitre~\cite{Lemaitre}, the related
discussion by Israel~\cite{Israel}, and more recently, the paper by Kraus
and Wilczek~\cite{Kraus-Wilczek}.  The \Painleve--Gullstrand
coordinates are related to the more usual Schwarzschild coordinates by
\begin{equation}
t_{PG} = t_{S} \pm 
\left[ 
4 M \; \hbox{\rm arctanh}\left(\sqrt{2GM\over r}\right) - 2 \; \sqrt{2GMr} 
\right].
\end{equation}
Or equivalently
\begin{equation}
dt_{PG} = dt_{S} \pm {\sqrt{2GM/r} \over 1-2GM/r}\; dr.
\end{equation}
With these explicit forms in hand, it becomes an easy exercise to
check the equivalence between the \Painleve--Gullstrand line element
and the more usual Schwarzschild form of the line element. It should
be noted that the $+$ sign corresponds to a coordinate patch that
covers the usual asymptotic region plus the region containing the
future singularity of the maximally extended Schwarzschild spacetime.
It thus covers the future horizon and the black hole singularity.  On
the other hand the $-$ sign corresponds to a coordinate patch that
covers the usual asymptotic region plus the region containing the past
singularity.  It thus covers the past horizon and the white hole
singularity.

As emphasized by Kraus and Wilczek, the \Painleve--Gullstrand line
element exhibits a number of features of pedagogical interest. In
particular the constant time spatial slices are completely flat ---
the curvature of space is zero, and all the spacetime curvature of the
Schwarzschild geometry has been pushed into the time--time and
time--space components of the metric.

Given the \Painleve--Gullstrand line element, it might seem trivial to
force the acoustic metric into this form: simply take $\rho$ and $c$
to be constants, and set $v=\sqrt{2GM/r}$? While this certainly forces
the acoustic metric into the \Painleve--Gullstrand form the problem
with this is that this assignment is incompatible with the continuity
equation $\vec\nabla\cdot(\rho\vec v)\neq 0$ that was used in deriving
the acoustic equations.

The best we can actually do is this: Pick the speed of sound $c$
to be a position independent constant, which we normalize to unity
($c=1$). Now set $v=\sqrt{2GM/r}$, and use the continuity equation
$\vec\nabla\cdot(\rho\vec v)= 0$ to deduce $\rho|\vec v| \propto
1/r^2$ so that $\rho \propto r^{-3/2}$. Since the speed of sound
is taken to be constant we can integrate the relation $c^2 =
dp/d\rho$ to deduce the equation of state must be $p = p_\infty +
c^2 \rho$ and that the background pressure satisfies $p-p_\infty
\propto c^2 r^{-3/2}$. Overall the acoustic metric is now
\begin{equation}
ds^2  \propto  r^{-3/2}
\left[ - dt^2 +
\left( dr \pm \sqrt{2GM \over r} dt \right)^2 
+ r^2\left( d\theta^2 + \sin^2\theta \; d\phi^2 \right) 
\right].
\end{equation}
The net result is conformal to the \Painleve--Gullstrand form of
the Schwarzschild geometry but not identical to it. For many purposes
this is quite good enough: we have an event horizon, we can define
surface gravity, we can analyze Hawking radiation. Since surface
gravity and Hawking temperature are conformal invariants~\cite{Jacobson93b}
this is sufficient for analyzing basic features of the Hawking
radiation process. The only way in which the conformal factor can
influence the Hawking radiation is through backscattering off the
acoustic metric.  (The phonons are minimally coupled scalars, not
conformally coupled scalars so there will in general be effects on
the frequency-dependent greybody factors.)

If we focus attention on the region near the event horizon, the
conformal factor can simply be taken to be a constant, and we can
ignore all these complications.

\section{The canonical acoustic black hole}

We can turn this argument around and ask: Given a spherically symmetric
flow of incompressible fluid, what is the acoustic metric? What is the
corresponding line element in Schwarzschild coordinates? If we start
by assuming incompressibility and spherical symmetry, then since
$\rho$ is position independent the continuity equation implies
$v\propto 1/r^2$. But if $\rho$ is position independent then (because
of the barotropic assumption) so is the pressure, and hence the speed
of sound as well. So we can define a normalization constant $r_0$ and
set
\begin{equation}
v = c \; {r_0^2\over r^2}.
\end{equation}
The acoustic metric is therefore, up to an irrelevant position-independent
factor,
\begin{equation}
ds^2 = 
- c^2 dt^2 + \left(dr \pm c \; {r_0^2\over r^2} \; dt \right)^2 
+ r^2 (d\theta^2 + \sin^2\theta \; d\phi^2).
\end{equation}
If we make the coordinate change
\begin{equation}
d\tau = dt \pm {r_0^2/r^2 \over c [1-(r_0^4/r^4)]} \; dr,
\end{equation}
then
\begin{equation}
ds^2 = 
- c^2  [1-(r_0^4/r^4)] d\tau^2 + {dr^2\over 1-(r_0^4/r^4)}
+ r^2 (d\theta^2 + \sin^2\theta \; d\phi^2).
\end{equation}
This is not any of the standard geometries typically considered in
general relativity but is, in the sense described above, the canonical
acoustic black hole.

It is very important to realise that a time-dependent version of
this canonical acoustic metric is very easy to set up
experimentally~\cite{Hochberg}, since the time-dependent version of
this canonical black hole metric is exactly the acoustic metric
that is set up around a spherically-symmetric bubble with oscillating
radius. For a bubble of radius $R$ we have
\begin{equation}
r_0 = R \sqrt{\dot R \over c}.
\end{equation}
We should only use this canonical metric for the fluid region outside
the bubble, and only in the approximation that the ambient fluid
is incompressible (e.g. water). For the typically gaseous and
necessarily compressible medium inside the bubble (e.g. air) we
should use a separate acoustic metric. The two acoustic metrics
need not be continuous across the bubble wall.

It is experimentally easy to generate (non-stationary) acoustic
apparent horizons in this manner: In cavitating bubbles (typically
air bubbles in water) it is experimentally easy to get the bubble
wall moving at supersonic speeds (up to Mach 10 in extreme cases).
Once the bubble wall is moving supersonically an acoustic apparent
horizon forms. It first forms at the bubble wall itself but then
will typically detach itself from the bubble wall (since the
apparent horizon will continue to be the surface at which the fluid
achieves Mach 1) as the bubble wall goes supersonic.  Since the
bubble must eventually stop its collapse and re-expand, there is
strictly speaking no acoustic event horizon (no absolute horizon)
in this experimental situation, merely a temporary apparent horizon.
(The apparent horizon must by construction last less than one
sound-crossing time for the collapsing bubble.)

To set up a geometry of this particular type with a true event
horizon (or at the very least, an apparent horizon that lasts for
many sound crossing times) requires a rather different physical
setup: a big tank of fluid with a long thin pipe leading to the
center. Then apply pressure to the tank till the outflow of fluid
escaping through the pipe goes supersonic, being careful to maintain
laminar flow and avoid turbulence. This would appear to be a
technologically challenging project.

\section{Hawking radiation and ``surface gravity''}

Establishing the existence of acoustic Hawking radiation follows
directly from the original Hawking argument~\cite{Hawking74,Hawking75}
once one realizes that the acoustic fluctuations effectively couple
to the Lorentzian acoustic metric introduced above. The only subtlety
arises in correctly identifying the ``surface gravity'' of an
acoustic black hole. Because of the definition of event horizon in
terms of phonons (null geodesics) that cannot escape the acoustic
black hole, the event horizon is automatically a null surface, and
the generators of the event horizon are automatically null geodesics.

In the case of acoustics there is one particular parameterization
of these null geodesics that  is ``most natural'', which is the
parameterization in terms of the Newtonian time coordinate of the
underlying physical metric.  This allows us to unambiguously define
a ``surface gravity'' even for non-stationary (time-dependent)
acoustic event horizons, by calculating the extent to which this
natural time parameter fails to be an affine parameter for the null
generators of the horizon. (This part of the construction fails in
general relativity where there is no universal natural time-coordinate
unless there is a timelike Killing vector --- this is why extending
the notion of surface gravity to non-stationary geometries in general
relativity is so difficult.)

When it comes to explicitly calculating the surface gravity in
terms of suitable gradients of the fluid flow, it is nevertheless
very useful to limit attention to situations of steady flow (so
that the acoustic metric is stationary). This has the added bonus
that for stationary geometries the notion of ``acoustic surface
gravity'' in acoustics is unambiguously equivalent to the general
relativity definition.

It is also useful to take cognizance of the fact that the situation
simplifies considerably for static (as opposed to merely stationary)
acoustic metrics.

\subsection{Static acoustic geometries}

To set up the appropriate framework, write the general stationary
acoustic metric in the form
\begin{equation}
ds^2 = {\rho\over c} 
\left[ - c^2 dt^2 + (d\vec x - \vec v \; dt)^2 \right].
\end{equation}
The time translation Killing vector is simply 
$ K^\mu = (1;\vec 0\,)$, with 
\begin{equation}
K^2 \equiv g_{\mu\nu} K^\mu K^\nu \equiv - ||K||^2 = -
{\rho\over c}[ c^2 - v^2]. 
\end{equation}
The metric can also be written as
\begin{equation}
ds^2 = {\rho\over c} 
\left[ 
-(c^2-v^2) dt^2  - 2 \vec v\cdot d\vec x \; dt + (d\vec x\,)^2 
\right].
\end{equation}
Now suppose that the vector $\vec v/(c^2 - v^2)$ is integrable,
then we can define a new time coordinate by
\begin{equation}
d\tau = dt + {\vec v \cdot d\vec x \over c^2 - v^2 }.
\end{equation}
Substituting this back into the acoustic line element
gives\footnote{The corresponding formula in~\cite{Unruh94} is missing
a factor of $c$ and a bracket.}
\begin{equation}
ds^2 = {\rho\over c} 
\left[ 
-(c^2-v^2) d\tau^2  + 
\left\{ \delta_{ij} + {v^i v^j \over c^2 - v^2} \right \}
dx^i dx^j
\right].
\end{equation}
In this coordinate system the absence of the time-space cross-terms
makes manifest that the acoustic geometry is in fact static (the
Killing vector is hypersurface orthogonal). The condition that an
acoustic geometry be static, rather than merely stationary, is thus
seen to be
\begin{equation}
\vec \nabla \times \left\{ {\vec v\over (c^2 - v^2)} \right\} = 0,
\end{equation}
that is
\begin{equation}
\vec v \times \vec \nabla (c^2 - v^2) = 0.
\end{equation}
This requires the fluid flow to be parallel to another vector that is
not quite the acceleration but is closely related to it. (Note that,
because of the vorticity free assumption, ${1\over2} \nabla v^2$ is
just the three-acceleration of the fluid, it is the occurrence of a
possibly position dependent speed of sound that complicates the
above.)

Once we have a static geometry, we can of course directly apply
all of the standard tricks~\cite{Membrane} for calculating the
surface gravity developed in general relativity. We set up a system
of fiducial observers (FIDOS) by properly normalizing the
time-translation Killing vector
\begin{equation}
V_{FIDO} \equiv 
{K\over ||K|| } = 
{K\over\sqrt{(\rho/c)[c^2-v^2]}}.
\end{equation}
The four-acceleration of the FIDOS is defined as $A_{FIDO} \equiv
(V_{FIDO} \cdot \nabla) V_{FIDO}$, and using the fact that $K$ is
a Killing vector, it may be computed in the standard manner
\begin{equation}
A_{FIDO} = +{1\over2} {\nabla ||K||^2 \over ||K||^2 }.
\end{equation}
That is 
\begin{equation}
A_{FIDO} = {1\over2} 
\left[ 
{\nabla (c^2-v^2) \over (c^2-v^2)}
+ {\nabla (\rho/c) \over (\rho/c)}
\right].
\end{equation}
The surface gravity is now defined by taking the norm $||A_{FIDO}||$,
multiplying by the lapse function, $ ||K|| = \sqrt{(\rho/c)[c^2-v^2]}$,
and taking the limit as one approaches the horizon: $|v|\to c$ ---
remember this is the static case. The net result is
\begin{equation}
||A_{FIDO}|| \; ||K|| = 
{1\over2} \vec v \cdot \nabla (c^2-v^2) + O(c^2-v^2),
\end{equation}
so that the surface gravity is given in terms of a normal derivative
by\footnote{Because of the background Minkowski metric there can be no
possible confusion as to the definition of this normal derivative.}
\begin{equation}
g_{H} = {1\over2} {\partial (c^2-v^2)\over\partial n} = 
 c \; {\partial (c-v) \over \partial n}.
\end{equation}
This is not quite Unruh's result~\cite{Unruh81,Unruh94} since he
implicitly took the speed of sound to be a position-independent
constant. The fact that $\rho$ drops out of the final result for
the surface gravity can be justified by appeal to the known conformal
invariance of the surface gravity~\cite{Jacobson93b}. Though derived
in a totally different manner, this result is also compatible with
the expression for ``surface-gravity'' obtained in the solid-state
black holes of Reznik~\cite{Reznik97}, wherein a position dependent
(and singular) refractive index plays a role analogous to the
acoustic metric. As a further consistency check, one can go to the
spherically symmetric case and check that this reproduces the
results for ``dirty black holes'' enunciated in~\cite{Visser92}.

Since this is a static geometry, the relationship between the
Hawking temperature and surface gravity may be verified in the
usual fast-track manner --- using the Wick rotation trick to
analytically continue to Euclidean space~\cite{Gibbons-Hawking}.
If you don't like Euclidean signature techniques (which are in any
case only applicable to equilibrium situations) you should go back
to the original Hawking derivations~\cite{Hawking74,Hawking75}.

One final comment to wrap up this section: the coordinate transform we
used to put the acoustic metric into the explicitly static form is
perfectly good mathematics, and from the general relativity point of
view is even a simplification. However, from the point of view of the
underlying Newtonian physics of the fluid, this is a rather bizarre
way of deliberately de-synchronizing your clocks to take a perfectly
reasonable region --- the boundary of the region of supersonic flow
--- and push it out to ``time'' plus infinity. From the fluid dynamics
point of view this coordinate transformation is correct but perverse,
and it is easier to keep a good grasp on the physics by staying with
the original Newtonian time coordinate.

\subsection{Stationary but non-static acoustic geometries}

If the fluid flow does not satisfy the integrability condition which
allows us to introduce an explicitly static coordinate system, then
defining the surface gravity is a little trickier. The situation is
somewhat worse than for general relativity since in the acoustic case
we have no reason to believe that anything like the zeroth law of
black hole mechanics holds~\cite{Laws}, nor do we have any reason to
believe that stationary event horizons have to be Killing horizons. 

Recall that the zeroth law of black hole mechanics (constancy of
the surface gravity over the event horizon) is proved in general
relativity by appealing to the Einstein equations and imposing
suitable energy conditions. In the acoustic paradigm we have no
analogue for the Einstein equations and no particular reason to
suspect the existence of anything like a zeroth law. Sufficiently
convoluted supersonic flows would seem to be able to set up almost
any pattern of surface gravity one wants.

Similarly, in general relativity the fact that stationary but
non-static black holes possess Killing horizons is related to the
axisymmetry that is deduced from the fact that non-axisymmetric
black holes are expected to lose energy via gravitational radiation
and so dynamically relax to an axisymmetric configuration --- in
the fluid dynamic models discussed here I have explicitly allowed
for external driving forces and explicitly excluded back reaction
effects, therefore there is no particular reason to expect acoustic
black holes to dynamically relax to axisymmetry.  In particular,
this means that even for stationary acoustic geometries there is
no particular reason to expect the acoustic event horizon to in
general be a Killing horizon.

So what does survive of our usual general relativistic notions for
acoustic event horizons in stationary but non-static geometries?
Recall that by construction the acoustic apparent horizon is in
general defined to be a two-surface for which the normal component
of the fluid velocity is everywhere equal to the local speed of
sound, whereas the acoustic event horizon is characterized by the
boundary of those null geodesics (phonons) that do  not escape to
infinity. In the stationary case these notions coincide, and it is
still true that the horizon is a null surface, and that the horizon
can be ruled by an appropriate set of null curves. Suppose we have
somehow isolated the location of the acoustic horizon, then in the
vicinity of the horizon we can split up the fluid flow into normal
and tangential components
\begin{equation}
\vec v = \vec v_\perp+ \vec v_\parallel; 
\qquad \hbox{where} \qquad
\vec v_\perp = v_\perp  \hat n .
\end{equation}
Here (and for the rest of this particular section) it is essential
that we use the natural Newtonian time coordinate inherited from
the background Newtonian physics of the fluid. In addition $\hat
n$ is a unit vector field that at the horizon is perpendicular to
it, and away from the horizon is some suitable smooth extension.
(For example, take the geodesic distance to the horizon and consider
its gradient.) We only need this decomposition to hold in some open
set encompassing the horizon and do not need to have a global
decomposition of this type available. Furthermore, by definition
we know that $v_\perp = c$ at the horizon. Now consider the vector
field
\begin{equation}
L^\mu = (1; v_\parallel^i).
\end{equation}
Since the spatial components of this vector field are by definition
tangent to a constant time slice through the horizon, the
integral curves of this vector field will be generators for the
horizon. Furthermore the norm of this vector (in the acoustic
metric) is
\begin{equation}
|| L ||^2 = 
-{\rho\over c} 
\left[ 
\vphantom{\Big|}
-(c^2 - v^2)  
- 2\vec v_\parallel \cdot \vec v 
+ \vec v_\parallel \cdot \vec v_\parallel 
\right]
\propto (c^2 - v_\perp^2).
\end{equation}
In particular, on the acoustic horizon $L^\mu $ defines a null
vector field, the integral curves of which are generators for the
acoustic horizon. I shall now verify that these generators are
geodesics, though the vector field $L$ is not normalized with an
affine parameter, and in this way shall calculate the surface
gravity. (For clarity, I will drop the conformal factor because I
already know that it will not affect the surface
gravity~\cite{Jacobson93b}.)

Consider the quantity $(L\cdot\nabla)L$ and calculate
\begin{equation}
L^\alpha \nabla_\alpha L^\mu 
= 
L^\alpha (\nabla_\alpha L_\beta - \nabla_\beta L_\alpha) g^{\beta \mu}
+ {1\over2} \nabla_\beta (L^2) g^{\beta \mu}.
\end{equation}
To calculate the first term note that
\begin{equation}
L_\mu = {\rho\over c} \; (-[c^2-v_\perp^2]; \vec v_\perp).
\end{equation}
Thus 
\begin{equation}
L_{[\alpha,\beta]} =
-\left[ 
\matrix{
0&\vdots&-\nabla_i\left[{\rho\over c}(c^2-v_\perp^2)\right]\cr
\cdots\cdots\cdots\cdots&\cdot&\cdots\cdots\cr
+\nabla_j\left[{\rho\over c}(c^2-v_\perp^2)\right]&\vdots&
\left({\rho\over c} \; v^\perp\right){}_{[i,j]}\cr 
} 
\right].	       
\end{equation}
And so:
\begin{equation}
L^\alpha L_{[\beta,\alpha]} = 
\left(
v_\parallel \cdot \nabla\left[{\rho\over c}(c^2-v_\perp^2)\right]; 
\nabla_j\left[{\rho\over c}(c^2-v_\perp^2)\right] 
+ v_\parallel^i \left({\rho\over c} \; v^\perp\right)_{[j,i]} 
\right).
\end{equation}
On the horizon, where $c=v_\perp$, this simplifies tremendously
\begin{equation}
(L^\alpha L_{[\beta,\alpha]})|_{horizon} = 
-{\rho\over c} \; (0; \nabla_j(c^2-v_\perp^2)).
\end{equation}
Similarly, for the second term we have
\begin{equation}
\nabla_\beta (L^2) = 
\left(0; \nabla_j\left[{\rho\over c}(c^2-v_\perp^2)\right] \right).
\end{equation}
On the horizon this again simplifies
\begin{equation}
\nabla_\beta (L^2)|_{horizon} = 
+{\rho\over c} \; (0; \nabla_j(c^2-v_\perp^2) ).
\end{equation}
There is partial cancellation between the two terms, and so
\begin{equation}
L^\alpha \nabla_\alpha L^\mu 
=  
+{1\over2 c^2} 
\left(v^j \nabla_j[(c^2-v_\perp^2)] ; 
   (c^2 \delta^{ij} - v^i v^j) \nabla_j[(c^2-v_\perp^2)] \right).
\end{equation}
But, as we have already seen, at the horizon the gradient term is
purely normal. Thus
\begin{equation}
L^\alpha \nabla_\alpha L^\mu 
=  +{1\over2 c} {\partial(c^2-v_\perp^2)\over \partial n} \;
    (1; v_\parallel^i). 
\end{equation}
Comparing this with the standard definition of surface
gravity~\cite{Wald}\footnote{There is an issue of normalization
here. On the one hand we want to be as close as possible to general
relativistic conventions. On the other hand, we would like the
surface gravity to really have the dimensions of an acceleration. The
convention adopted here is the best compromise I have come up with.}
\begin{equation}
L^\alpha \nabla_\alpha L^\mu 
=  + {g_H\over c} L^\mu. 
\end{equation}
we finally have
\begin{equation}
g_H = 
{1\over2} {\partial(c^2-v_\perp^2)\over \partial n} =
 c \; {\partial(c-v_\perp)\over\partial n}.
\end{equation}
This is in agreement with the previous calculation for static
acoustic black holes, and insofar as there is overlap, is also
consistent with results of Unruh~\cite{Unruh81,Unruh94},
Reznik~\cite{Reznik97}, and  the results for ``dirty black
holes''~\cite{Visser92}. From the construction it is clear that
the surface gravity is a measure of the extent to which the Newtonian
time parameter inherited from the underlying fluid dynamics fails
to be an affine parameter for the null geodesics on the horizon.

\section{Geometric Acoustics}

Up to now, have been developing general machinery to force acoustics
into Lorentzian form. This can be justified either with a view to using
fluid mechanics to teach us more about general relativity, or to using
the techniques of Lorentzian geometry to teach us more about fluid
mechanics.

For example, given the machinery developed so far, taking the short
wavelength/high frequency limit to obtain geometrical acoustics is now
easy. Sound rays (phonons) follow the {\em null geodesics} of the
acoustic metric. Compare this to general relativity where in the
geometrical optics approximation light rays (photons) follow {\em null
geodesics} of the physical spacetime metric. Since null geodesics are
insensitive to any overall conformal factor in the
metric~\cite{MTW,Hawking-Ellis,Wald} one might as well simplify life by
considering a modified conformally related metric
\begin{equation}
h_{\mu\nu} \equiv 
\left[ \matrix{-(c^2-v_0^2)&\vdots&-v_0^j\cr
               \cdots\cdots\cdots\cdots&\cdot&\cdots\cdots\cr
	       -v_0^i&\vdots&\delta^{ij}\cr } 
\right]_.	       
\end{equation}
This immediately implies that, in the geometric acoustics limit,
sound propagation is insensitive to the density of the fluid. In
this limit, acoustic propagation depends only on the local speed
of sound and the velocity of the fluid. It is only for specifically
wave related properties that the density of the medium becomes
important.

We can rephrase this in a language more familiar to the acoustics
community by invoking the Eikonal approximation. Express the
linearized velocity potential, $\psi_1$, in terms of an amplitude,
$a$, and phase, $\varphi$, by $\psi_1 \sim a e^{i\varphi}$. Then,
neglecting variations in the amplitude $a$, the wave equation
reduces to the {\em Eikonal equation}
\begin{equation}
h^{\mu\nu} \; \partial_\mu \varphi \; \partial_\nu \varphi = 0.
\end{equation}
This Eikonal equation is blatantly insensitive to any overall
multiplicative prefactor (conformal factor).

As a sanity check on the formalism, it is useful to re-derive some
standard results.  For example, let the null geodesic be parameterized
by $x^\mu(t) \equiv (t; \vec x(t))$. Then the null condition implies
\begin{eqnarray}
&& h_{\mu\nu} {dx^\mu \over dt} {dx^\nu \over dt} = 0
\nonumber\\
&&\iff
-(c^2 - v_0^2) - 2 v_0^i {dx^i \over dt} 
+ {dx^i \over dt} {dx^i \over dt} = 0
\nonumber\\
&&\iff
\left\Vert {d{\vec x} \over dt} - \vec v_0 \right\Vert = c.
\end{eqnarray}
Here the norm is taken in the flat physical metric. This has the
obvious interpretation that the ray travels at the speed of sound,
$c$, relative to the moving medium.

Furthermore, if the geometry is stationary one can do slightly
better. Let $x^\mu(s) \equiv (t(s); \vec x(s))$ be some null path
from $\vec x_1$ to $\vec x_2$, parameterized in terms of physical
arc length ({\em i.e.} $|| d{\vec x}/ds || \equiv 1$). Then the
tangent vector to the path is
\begin{equation}
{dx^\mu\over ds} = \left( {dt\over ds}; {dx^i\over ds} \right).
\end{equation}
The condition for the path to be null (though not yet necessarily
a null geodesic) is
\begin{equation}
g_{\mu\nu} {dx^\mu\over ds} {dx^\nu\over ds} = 0.
\end{equation}
Using the explicit algebraic form for the metric, this can be
expanded to show
\begin{equation}
-(c^2 - v_0^2) \left({dt\over ds}\right)^2 
- 2 v_0^i \left({dx^i\over ds}\right) \left({dt\over ds}\right)
+1 = 0.
\end{equation}
Solving this quadratic 
\begin{equation}
\left({dt\over ds}\right) 
= { - v_0^i \left({dx^i\over ds}\right)
    + \sqrt{ c^2 - v_0^2 + \left(v_0^i \; {dx^i\over ds}\right)^2 }
   \over
   c^2 - v_0^2}.
\end{equation}
Therefore, the total time taken to traverse the path is  
\begin{eqnarray}
T[\gamma] 
&=& \int_{\vec x_1}^{\vec x_2} (dt/ds) \; ds 
\nonumber\\
&=& \int_\gamma {1\over c^2-v_0^2} 
                 \Big\{  
                 \sqrt{ (c^2 - v_0^2)ds^2 + (v_0^i \; dx^i)^2 } 
		 - v_0^i \; dx^i 
                 \Big\}.
\end{eqnarray}
If we now recall that extremizing the total time taken is Fermat's
principle for sound rays, we see that we have checked the formalism
for stationary geometries (steady flow) by reproducing the discussion
on page 262 of Landau and Lifshitz~\cite{Landau-Lifshitz}.

As a second example of the insights arising from the Lorentzian point of
view consider the ``reciprocity theorem''. Suppose a pulse of sound
is emitted at time $t_1$ at position $\vec x_1$. The disturbance
propagates according to the inhomogeneous differential equation
\begin{equation}
\Delta \psi = 
{1\over\sqrt{-g}} \delta^4(x^\mu - x^\mu_1) = 
{c\over\rho^2} \; \delta(t-t_1) \; \delta^3(\vec x - \vec x_1).
\end{equation}
The solution to this is the retarded scalar Green function
\begin{equation}
\psi(x)|_{{source~at~x_1}} = G_R(x,x_1).
\end{equation}
The Green function has well known symmetry properties that are
completely unaffected by any time dependence in the underlying
acoustic metric. We may in the usual manner, construct advanced
and retarded Green functions that vanish outside the past and future
sound cones respectively. Then
\begin{equation}
G_R(x_2,x_1) = G_A(x_1,x_2).
\end{equation}
So that the reciprocity theorem {\em for the velocity potential}
is valid in absolute generality. 
\begin{equation}
\psi(x_2)|_{{source~at~x_1}} = 
\psi(x_1)|_{{source~at~x_2}}^{time~reversed}.
\end{equation}
To get a reciprocity theorem for {\em pressure} one has to recall
\begin{equation}
p_1 =  \rho_0 ( \partial_t \psi_1 + \vec v_0 \cdot \nabla\psi_1 ).
\end{equation}
Then, by restricting to the case of fluid at rest ($\vec v_0=0$,
$\partial_t \rho_0=0$, $\partial_t p_0=0$), using the time translation
invariance of the Green functions, and the time reversal property $G_A
\rightleftharpoons G_R$, one has
\begin{equation}
\left[{p_1\over\rho_0}\right](\vec x_2, t_2-t_1)
|_{{source~at~x_1}} =
\left[{p_1\over\rho_0}\right](\vec x_1, t_2-t_1)
|_{{source~at~x_2}}.
\end{equation}
This result is still much more general than the usual reciprocity theorem.

\section{Limitations}

The derivation of the wave equation made two key assumptions ---
that the flow is irrotational flow and the fluid is barotropic.

The d'Alembertian equation of motion for acoustic disturbances,
though derived only under the assumption of irrotational
flow,\footnote{But remember that irrotational flow is automatic for
superfluids~\cite{Comer}, and is natural in situations of high
symmetry.} continues to make perfectly good sense in its own right
if the background velocity field $\vec v_0$ is given some vorticity.
This leads one to hope that it {\em might} be possible to find a
suitable generalization of the present derivation that might work
for flows with nonzero vorticity. In this regard, note that if the
vorticity is everywhere confined to thin vortex filaments, the
present derivation already works everywhere outside the vortex
filaments themselves.

The technical problem with flows with non-zero vorticity is that
the vorticity in the background flow couples to the perturbations
and generates vorticity in the fluctuations. Then sound waves can
no longer be represented simply by a scalar potential and a much
more complicated mathematical structure results. (Phonons are no
longer simply minimally coupled scalar fields and the appropriate
generalization is sufficiently unpleasant as to be intractable.)

The restriction to a barotropic fluid ($\rho$ a function of $p$
only) is in fact also related to issues of vorticity. Examples of
barotropic fluids are:
\begin{itemize}
\item 
Isothermal fluids subject to isothermal perturbations.
\item 
Fluids in convective equilibrium subject to adiabatic perturbations.
\end{itemize}
See for example \cite{Lamb}, \S 311, pp 547--548, and \S 313 pp
554--556.  Failure of the barotropic condition implies that the
perturbations cannot be vorticity free and thus requires more
sophisticated analysis.

If the fluid is in addition inviscid then the analysis of this paper
implies a hidden Lorentz invariance in the acoustic equations. This
hidden Lorentz invariance is more than just a formal quirk: If one has
``atoms'' held together by phonons (Cooper pairs?), then
these atoms, and complex systems built up out of such atoms, will see
(hear) an acoustic special relativity that is as real to them as
Einstein's special relativity is to us.  Furthermore these systems
would with additional observation detect (hear) an acoustic general
relativity --- but instead of the Einstein--Hilbert equations of our
general relativity they would experience an acoustic general
relativity governed by the hydrodynamic equations.

If the fluid has non-zero viscosity then there will be violations of
this acoustic Lorentz symmetry. These violations are momentum dependent
and, as I shall discuss in the next section, they are small at low
momentum.

\section{Viscosity: breaking the Lorentz symmetry}

After this long build-up emphasizing the hidden Lorentzian geometry
hiding in (inviscid vorticity--free barotropic) fluid dynamical
equations, I will now show how to explicitly break the Lorentz
symmetry.  From the atomic perspective underlying continuum fluid
mechanics the eventual breakdown of the Lorentz symmetry governing
the notion of the phonons is no great surprise: eventually, once
the wavelength of the phonons is less than the mean interatomic
spacing in the fluid, we should certainly expect modifications to
the phonon dispersion relation~\cite{Unruh81,Jacobson91}.  Specific
{\em ad hoc} mutilations of the dispersion relation have been
considered by Jacobson~\cite{Jacobson91}, Unruh~\cite{Unruh94},
Corley and Jacobson~\cite{Corley-Jacobson96}, and Corley~\cite{Corley97a}.
I shall now show that a similar but not identical breakdown of
acoustic Lorentz invariance can be deduced directly from the
continuum equations merely by adding the effects of viscosity.

Of course, the fundamental equations of fluid dynamics, the equation
of continuity (\ref{E-continuity}) and Euler's equation (\ref{E-euler})
are unaltered. What changes is the expression for the driving force
in Euler's equation so that (\ref{E-force}) becomes~\cite[\S 328,
page 576-577]{Lamb}
\begin{equation}
\vec F = - \nabla p - \rho \nabla \phi - \rho \nabla\Phi + 
\rho \; \nu 
\left(\nabla^2 \vec v + {1\over3} \vec\nabla (\vec\nabla\cdot\vec v) \right).
\label{E-force-nu}
\end{equation}
Here $\nu$ denotes kinematic viscosity.  I again take the flow to
be {\em vorticity free}, and again choose the fluid to be {\em barotropic}.
Repeating the steps that led to (\ref{E-bernoulli}) now show that
Euler's equation reduces to
\begin{equation}
-\partial_t \psi + h + {{1\over2}} (\nabla\psi)^2 
+ \phi  + \Phi + {4\over3} \nu \; \nabla^2 \psi = 0.
\label{E-burgers}
\end{equation}
This again is a well-known equation, simply being Burgers' equation
subject to external driving forces~\cite{Burgers}. (In obtaining
this equation it is necessary to assume that the kinematic viscosity
$\nu$ is position independent. In addition it is common practice,
though not universal, to absorb the $4/3$ into a modified definition
of kinematic viscosity.)

Linearization proceeds as previously. For the continuity equation
there are no changes, while linearizing the Euler equation (Burgers'
equation) yields
\begin{eqnarray}
&&-\partial_t \psi_0 + h_0 + {{1\over2}} (\nabla\psi_0)^2 
+ \phi + \Phi + {4\over3} \nu \; \nabla^2 \psi_0= 0.
\\
&&-\partial_t \psi_1 + {p_1\over\rho_0} - \vec v_0 \cdot \nabla\psi_1  
+{4\over3} \nu \; \nabla^2 \psi_1= 0.
\end{eqnarray}
Rearranging
\begin{equation}
p_1 =  \rho_0
\left( 
\partial_t \psi_1 + \vec v_0 \cdot \nabla\psi_1 
- {4\over3} \nu \nabla^2 \psi_1
\right).
\label{E-linear-euler-nu}
\end{equation}
As before, we substitute this linearized Euler equation into the
linearized continuity equation, to obtain the physical wave equation:
\begin{eqnarray}
&-& 
\partial_t  
\left( 
{\partial\rho\over\partial p} \; \rho_0 \; 
\left( 
\partial_t \psi_1 + \vec v_0 \cdot \nabla\psi_1 - 
{4\over3} \nu \;\nabla^2 \psi_1
\right) 
\right)
\nonumber\\
&+& 
\nabla \cdot 
\left( 
\rho_0 \; \nabla\psi_1 
- {\partial\rho\over\partial p} \; \rho_0 \; \vec v_0 \;
\left( 
\partial_t \psi_1 + \vec v_0 \cdot \nabla\psi_1 - 
{4\over3} \nu \;\nabla^2 \psi_1
\right) 	    
\right)
=0.
\label{E-wave-physical-nu}
\end{eqnarray}
Using the same matrix $f^{\mu\nu}$ defined previously the above wave
equation is easily rewritten as\footnote{I have used the continuity
equation for the background fluid flow to pull the factor $\rho_0$
outside the convective derivative.}
\begin{equation}
\partial_\mu ( f^{\mu\nu} \; \partial_\nu \psi_1) =
 - {4\over3} \rho_0 \; \nu \;
\left({\partial\over\partial t} + \vec v_0\cdot \vec \nabla\right)
\left[ c^{-2} \nabla^2 \psi_1 \right].
\end{equation}
In terms of the d'Alembertian associated with the acoustic metric
this reads
\begin{equation}
\Delta \psi_1 =
 - {4\over3} {\nu c\over\rho_0}
\left({\partial\over\partial t} + \vec v_0\cdot \vec \nabla\right)
\left[ c^{-2} \nabla^2 \psi_1 \right].
\end{equation}
The convective derivative appearing here may
easily be converted into four-dimensional form by utilizing the
acoustic four-velocity for the fluid. Recall that
\begin{equation}
V^\mu = { (1; \vec v\,) \over \sqrt{\rho_0 c} }.
\end{equation}
It is easy to see that this is a timelike unit vector in the acoustic
metric, so that
\begin{equation}
\Delta \psi_1 =
 - {4\over3} {\nu c^2\over\sqrt{\rho_0 c}}
\left(V^\mu \nabla_\mu \right)
\left[ c^{-2} \nabla^2 \psi_1 \right].
\end{equation}
The $\nabla^2 \psi_1$ term explicitly couples only to the flat
spatial metric and can be written in terms of the acoustic metric
by noting that
\begin{equation}
g^{\mu\nu} = - V^\mu V^\nu + {c\over\rho} \; {}^{(3)}g_{space}^{\mu\nu}. 
\end{equation}
It is the explicit appearance of the fluid four-velocity in the
above expressions that justifies my claim that viscosity breaks
the acoustic Lorentz invariance.

\paragraph{Sanity check I:} 
If the background fluid flow is at rest and homogeneous ($\vec v_0
= 0$, and with $\rho_0$ and $c$ independent of position) then this
viscous wave equation reduces to
\begin{equation}
\partial_t^2  \psi_1  = c^2 \nabla^2 \psi_1 
+ {4\over3} \nu \; \partial_t \nabla^2 \psi_1.
\label{E-lamb}
\end{equation}
This equation may be found, for instance, in \S 359 pages 646--648
of Lamb~\cite{Lamb}.

\paragraph{Sanity check II:} 
Take the Eikonal approximation in the form 
\begin{equation}
\psi_1 = a(x)
\exp(-i[\omega t - \vec k\cdot \vec x]\,),
\end{equation}
with $a(x)$ a slowly varying function of position. Furthermore, agree
to ignore derivatives of the metric. Then the viscous wave equation in
the Eikonal approximation reduces to
\begin{equation}
-(\omega - \vec v\cdot \vec k)^2 + c^2 k^2 
- i\nu \, {4\over3} (\omega - \vec v\cdot \vec k) k^2 = 0.
\end{equation}
This lets us write down a dispersion relation for sound waves
\begin{equation}
\omega =  
\vec v\cdot \vec k \pm 
\sqrt{c^2 k^2 - \left({2\nu k^2\over3}\right)^2 } 
- i {2\nu k^2\over3}.
\end{equation}
The first term simply arises from the bulk motion of the fluid.
The second term specifically introduces dispersion due to viscosity,
while the third term is specifically dissipative. The {\em ad hoc}
models introduced in Jacobson~\cite{Jacobson91}, Unruh~\cite{Unruh94},
Corley and Jacobson~\cite{Corley-Jacobson96}, and Corley~\cite{Corley97a}
are exactly recovered by ignoring the dissipation due to viscosity
but retaining the dispersion due to viscosity.

Note that the violation of Lorentz invariance is suppressed at low
momentum.  This is in agreement with general arguments of Nielsen
{\em et al}~\cite{Nielsen78,Nielsen83a,Nielsen83b}, though it should
be borne in mind that Nielsen {\em et al} were dealing with
interacting quantum field theories and the context here is, if not
purely classical, at worst one of free phonons propagating on a
fixed classical background.  (An alternative model for the breakdown
of Lorentz invariance has been discussed by
Everett~\cite{Everett76a,Everett76b}.) The violations of Lorentz
symmetry become significant once
\begin{equation}
k \approx k_0 \equiv {c\over \nu}.
\end{equation}
But from the atomic theory of (normal) fluids
\begin{equation}
\nu 
\approx {\hbox{mean free path}^2\over\hbox{mean free time} } 
\approx c \cdot \hbox{mean free path}.
\end{equation}
This gives the very reasonable result
\begin{equation}
k_0 \approx {1\over\hbox{mean free path}},
\end{equation}
verifying that the breakdown of acoustic Lorentz invariance is
explicitly linked to the atomic nature of matter.

\section{Precursors}

It is perhaps surprising that anything new can be said about so
venerable a subject as fluid dynamics. Certainly there are precursors
to the discussion of this paper in the fluid dynamics literature.
For instance, take the background to be static, so that $\vec v_0=0$,
while $\partial_t \rho_0 = 0 = \partial_t p_0$, though $p_0$ and hence
$c$ are permitted to retain arbitrary spatial dependencies. Then the
wave equation derived in this paper reduces to
\begin{equation}
\partial_t^2 \psi = c^2 {1\over \rho_0} \nabla \cdot (\rho_0 \nabla \psi).
\end{equation}
This equation is in fact well known. It is equivalent, for instance
to eq.~(13) of \S 313 of Lamb's classic {\em Hydrodynamics}~\cite{Lamb}.
See also eq.~(1.4.5) of the recent book by DeSanto~\cite{deSanto}.
The superficially similar wave equations discussed by Landau and
Lifshitz~\cite{Landau-Lifshitz} (see \S 74, eq.~(74.1)), and by
Skudrzyk~\cite{Skudrzyk} (see p 282), utilize somewhat different
physical assumptions concerning the behaviour of the fluid.

In a somewhat different vein, the modern study of classical continuum
mechanics has greatly benefited from the use of $3$--dimensional
Riemannian geometry to describe the physics of the spatial
configurations of elastic media and other
continua~\cite{Handbuch,Eringen,Leigh}. Analyses of this type have
traditionally treated space and time on quite separate footings.

The most direct precursor of the results derived in this paper are
due to Unruh~\cite{Unruh81} and Jacobson~\cite{Jacobson91}, and in
the body of work prompted by those
papers~\cite{Jacobson93,Unruh94,Brout,Jacobson95,Jacobson96,%
Corley-Jacobson96,Corley-Jacobson97,Corley97a,Corley97b,Reznik96,Reznik97}.

\section{Summary and Discussion}

Acoustic waves in an inviscid fluid can, under the assumptions of
irrotational barotropic flow, be described by an equation of motion
involving the scalar d'Alembertian of a suitable Lorentzian geometry.
For inhomogeneous flows this Lorentzian geometry will exhibit
nonzero Riemann curvature.

Traditionally, Lorentzian geometries have been of interest to
physics only within the confines of Einstein's theory of gravitation.
The results of this paper provides the general relativity community
with a very down to earth physical model for certain classes of
Lorentzian geometry. This is of interest both pedagogically and
because it extends the usefulness of Lorentzian differential geometry
beyond the confines of Einstein gravity.

Particularly intriguing is the fact that while the underlying
physics of fluid dynamics is completely nonrelativistic, Newtonian,
and sharply separates the notions of space and time, one nevertheless
sees that the acoustic fluctuations couple to a full--fledged
Lorentzian {\em spacetime}.

As discussed by Unruh~\cite{Unruh81}, (and subsequent
papers~\cite{Jacobson91,Jacobson93,Unruh94,Brout,Jacobson95,Jacobson96,%
Corley-Jacobson96,Corley-Jacobson97,Corley97a,Corley97b,Reznik96,Reznik97})
an acoustic event horizon will emit Hawking radiation in the form of a
thermal bath of phonons at a temperature
\begin{equation}
k T_H = {\hbar \; g_H\over 2\pi c}.
\end{equation}
(Yes, this really is the speed of sound in the above equation, and
$g_H$ is really normalized to have the dimensions of a physical
acceleration.) Using the numerical expression
\begin{equation}
T_H = 
(1.2\times 10^{-9} K m) \;
\left[ {c\over 1000 m s^{-1}} \right]\; 
\left[ {1\over c} {\partial(c-v_\perp)\over\partial n} \right],
\end{equation}
it is clear that experimental verification of this acoustic Hawking
effect will be rather difficult. (Though, as Unruh has pointed
out~\cite{Unruh81}, this is certainly technologically easier than
building [general relativistic] micro-black holes in the laboratory.)

A particularly important side effect of this entire analysis is
that it forces us to re-examine all of black hole physics to cleanly
separate what is intrinsic to general relativity from what is
generic to Lorentzian geometries. The acoustic analog for black
hole physics accurately reflects half of general relativity ---
the kinematics due to the fact that general relativity takes place
in a Lorentzian spacetime.  The aspect of general relativity that
does not carry over to the acoustic model is the dynamics --- the
Einstein equations. Thus the acoustic model provides a very concrete
and specific model for separating the kinematic aspects of general
relativity from the dynamic aspects.

In particular, perhaps the most important lesson to be learned is
this: Hawking radiation from event horizons is a purely kinematic
effect that occurs in any Lorentzian geometry with an event horizon
and is independent of any dynamical equations imposed on the
Lorentzian geometry. On the other hand, the classical laws of black
hole mechanics~\cite{Laws} are intrinsically results of the dynamical
equations (Einstein equations) that have no analog in the acoustic
model.  Thus Hawking radiation persists even in the absence of the
laws of black hole mechanics and, in particular, the existence or
otherwise of Hawking radiation is now seen to be divorced from the
issue of the existence or otherwise of the {\em laws of black hole
thermodynamics}.  Hawking radiation is a purely kinematical effect
that will be there regardless of whether or not it makes any sense
to assign an entropy to the event horizon --- and attempts at
deriving black hole entropy from the Hawking radiation phenomenon
are thereby seen to require specific dynamical assumptions about
the (at least approximate) applicability of the Einstein equations.

\section*{Acknowledgements}

This work was supported by the U.S. Department of Energy.  I
particularly wish to thank Ted Jacobson for encouraging me to
resuscitate this paper, and expand it into its current form. I also
wish to thank John Friedman and Ted Jacobson for bringing the Unruh
reference~\cite{Unruh81} to my attention when this work was in its
preliminary form~\cite{Visser93}. Additionally I wish to thank Greg
Comer~\cite{Comer} and David Hochberg~\cite{Hochberg} for kindly
providing me with access to unpublished manuscripts.


\end{document}